\documentclass[preprint,aps,showpacs]{revtex4}
\usepackage{epsfig}
\usepackage{graphicx}
\usepackage{amsmath}

\def\be{\begin{equation}}
\def\ee{\end{equation}}
\def\ba{\begin{eqnarray}}
\def\ea{\end{eqnarray}}

\begin{document}

\title
{Possible propagation of the Zhang-Rice singlet as a probable Cooper channel in the $CuO_2$ planes}

\author{G. E. Akpojotor}
\affiliation{Max-Planck-Institut f\"ur Physik komplexer Systeme, N\"othnitzer Strasse 38, 01187 Dresden, Germany}
\date{\today}

\begin{abstract}
The issue of how superconductivity originate in the $CuO_2$ planes believed to be crucial to understanding the high $T_c$ superconducting cuprates is still an going debate. In the wake of recent experimental observations of the   the Zhang-Rice singlet (ZRS), its formation and propagation need to be revisited especially by using a simple approach almost at a phenomenological level. Within a highly simplified correlated variational approach (HSCVA) in this paper, a new formation  of the ZRS as constituting the ground state of a single-band t-J model of the $CuO_2$ planes is developed. This formation is then used to demonstrate how the ZRS can be propagated as a probable Cooper channel in the $CuO_2$ planes.

\end{abstract}
\pacs{74.72.-h, 74.25.Jb, 74.20.-z}
\maketitle

\section {Introduction}
One early consensus after the discovery of the high $T_c$ superconducting cuprates, is that the key to understanding these materials is the $CuO_2$ planes common to all of them. However, after two decades of intense study, there is no consensus  understanding yet of the origin of superconductivity from this feature. Interestingly, it was experimentally demonstrated early enough that when the parent material is doped, the mobile holes reside on the O site of this plane. This led to the suggestion that a three-band Hubbard model ($H_{3b}$) in which the hole is mobile and carry a spin should be the starting point to investigate these material \cite{Emery1987}. In a seminal paper \cite{Zhang1988}, Zhang and Rice suggested that this hole on the O site will form a singlet state with a hole on the Cu site at the central of each $CuO_2$ plaquette to form a single-band character. This is the mapping of the $H_{3b}$ into the single-band Hubbard model ($H_{1b}$) when $J = 4t^2/U$ which is consistent with the Anderson's proposal that a strong on-site Coulomb interaction among a partially filled band of Cu 3d levels should be the starting effective single-band model for the supeconducting cuprates. Since then, the researchers who follow this line of thinking \cite{Belinicher1993, Gavrichkov2000} believe the ground states of the $CuO_2$ are the Zhang-Rice singlet (ZRS) which are expected to become the Cooper pairs of the superconducting states when liberated from the insulating host material. However, there have been opposing views on the equivalence of the $H_{3b}$ and $H_{1b}$ and also that the added holes do not hybridize into a ZRS \cite{Emery1988, Thomale2008}. These views may need some reconsideration in the wake of recent experiments \cite{Tjeng1997, Learmonth 2007, Moreshini2008} and calculations \cite{Belinicher1996, Hozoi2007, Yin2008}  which have demonstrated the existence of the ZRS. Meanwhile among the workers starting from the ZRS, there is still no consensus on the appropriate parameters to add to the t-J model  to obtain an effective Hamiltonian for the superconducting cuprates \cite{Yushankhai1997, Ogata2008}. The reason is that the effects of these parameters even in first principle calculations, depend on the starting $Cu0_2$ configuration. It follows then that a simplified determination of the possible configuration at a phenomenological level but not bias to experimental results can give useful insight into the starting structure of the $CuO_2$. This is the purpose of this paper.\\
Interestingly the ZRS is obviously a two-hole state \cite{Gavrichkov2000, Durr2000, Brugger 2008} and this makes it straightforward for us to apply to the $CuO_2$ planes our recently developed formulation of the statistical equivalent of the Hubbard (t-U and $t-t'-U$) model using a highly simplified correlated variational approach (HSCVA)\cite{Akpojotor2008}. The simplicity of this approach makes it easy to give useful insight on how the competition between itineracy and localization in strongly correlated systems could lead to the exciting properties in these systems in $d = 1, 2, 3$. In the present formulation, it will be shown how the competition between the motion of a single-hole in the the $CuO_2$ planes and its hybridization with the Cu can lead to a $H_{1b}$ model. In particular, the approach emphasizes the ZRS as the ground states of the $CuO_2$ planes and that it can be propagated as a probable Cooper channel in the $CuO_2$ structure within the t-J Hamiltonian for the supeconducting cuprates.
\section{Method and formulation}
It is generally believed that the undoped superconducting cuprates is a Mott insulator at half-filling (one electron per site). In this state, the system is more stable because the electron are reluctant to hop in order to avoid the price of occupying the same site (i.e U). Thus the starting $CuO_2$ is viewed as a valence state of $Cu\ 3d^{9}$ and $O\ 2P^6$ with a hole at the Cu site. One way to induce electronic motion is to create an electronic vacancy on the O site which is often called a hole. This free hole attracts an electron of another O which in turn creates a vacancy  in its place and this new vacancy will attract another electron and so on. Thus by this first electron removal (FER) approach, the electron can hop from one O site to another, but in doing so, the hole is also hopping from one point to another and this is often referred to as the hole motion. As stated above, the formulation in \cite{Zhang1988} assumes that the ZRS is formed from the hybridization of a hole on each square of O atoms to the central Cu to form the local singlet. It follows that this FER approach to investigate the propagation of a single hole in the $CuO_2$ plane is a natural starting point to investigate the superconducting cuprates and the ZRS states are believed to be the theoretical construction of the FER states in photoemission experiments \cite{Durr2000}.\\
In the theoretical construction here, we start from the $H_{3b}$ scenario by assuming all four possible Cu-O hybridized states in a plaquette. Since only one hole is present in the FER approach, it is argued that only one of the hybridized states will be present at any given time to form an effective singlet of each plaquette. Therefore this effective singlet retain the ZRS nature of the $CuO_2$ structure in the FER approach. The non-triviality of this consideration to the formation here will be made clear below.\\
\begin{figure}[htbp]
 {\includegraphics[width=8cm, height=10cm]{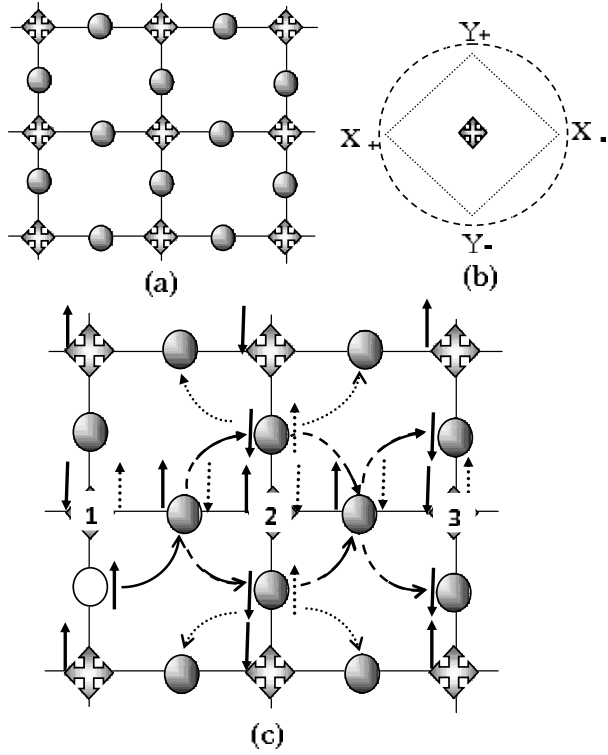}}
 \caption{(a) The 2D lattice of Cu-O layer of hight $T_c$ cuprtates showing the $CuO_2$ planes (b) The $CuO_2$ plaquette as a 1D ring of O around the Cu so that there is no edge effect hence have periodic boundaries (c) Hole doping leading to propagation of the ZRS in the $CuO_2$ planes.}
\label{fig:1}
\end{figure}
Following the steps in \cite{Akpojotor2008}, the basis states of the $CuO_2$ plaquatte in a three-band configuration from Fig. 1b are , $\vert Cu_x\sigma,O_{x_1}\bar{\sigma}>$, $\vert Cu_x\sigma,O_{x_2}\bar{\sigma}>$, $\vert Cu_y\sigma,O_{y_1}\bar{\sigma}>$, $\vert Cu_y\sigma,O_{y_2}\bar{\sigma}>$, where the $\sigma$ and $\bar{\sigma}$ represent the spins defined by  $\sigma(\bar{\sigma}) =  \uparrow(\downarrow),\downarrow(\uparrow)$. In numerical index, a state can be represented  in general with the first site for Cu and the second site for O as $\vert ii\sigma, jk\bar{\sigma}>$ where (ii) means there is a particle at the site i that can possibly be in the (x,y) direction.  Then the state $\vert Cu_x\sigma,O_{x_1}\bar{\sigma}>$ becomes $\vert i0\sigma, i_10\bar{\sigma}>$ where i = 0, 1, 2, 3, ...N is the index number of Cu sites. It is worthy to note that the notation used here in the configuration of the hybridized states is for convenience of the approach and do not differ from the common ones in the literature. Further more, this configuration combined with the periodicity of the $CuO_2$ structure implies the eigenstates of the $CuO_2$ plane are eigenstates of the translational operator, $\hat{T_r}\vert\psi> = exp(l\vec{k}.\vec{r})\vert\psi>$, and this gives a recipe to  construct translationally invariant basis states with additional fixed quantum number, $K =\frac{2\pi i}{N}$ which will become important to extend the simple approach here in future communication.\\
The $CuO_2$ plane is a quasi two-dimensional (2D) structure and therefore requires the use of the 2D kinetic operator, $h_t$, which for a general state, $\vert ii\sigma, jk\bar{\sigma}>$ say, yields the excited states simply by adding and subtracting one appropriately \cite{Akpojotor2008},
\begin{eqnarray}
&&h_t\vert ii\sigma, jk\bar{\sigma}>= -t[\vert (i \pm 1)i\sigma, jk\bar{\sigma}> + \vert i(i\pm 1)\sigma, jk\bar{\sigma}> + \nonumber\\
&&\vert ii\sigma, (j \pm 1)k\bar{\sigma}> + \vert ii\sigma, j(k \pm 1)\bar{\sigma}>],
\end{eqnarray}
where -t is the nearest neigbour (NN) hopping matrix.\\
It has been shown \cite{Akpojotor2008} that the variational ground state energy
\begin{equation}
 E_g = \frac{\langle\psi\vert H \vert\psi\rangle}{\langle\psi\vert \psi\rangle}
\end{equation}
for the model Hamiltonian H = $h_t$ (i.e Eq. 1) will yield a matrix
\begin{equation}
[E_\delta{_{{L'}_c L}} + (T)_{{L}_c L'}]=0
\end{equation}
where $E = E_g/t$ is the total ground state energy spectrum which can be obtained by diagonalization of Eq. (3) and its smallest value is the ground state energy of the systems. The $L_c$ in Eq. (3) is the separation between the sites in a basis state written in compact form, $L_c = \vert i - j \vert$, $L'$ is the new separation of the excited state obtained and T is the number of such excited states with $L'$ for an operation on any of the basis states. We emphasize that since the Cu-O distances in the plane are the same \cite {Gavrichkov2000}, the separation depends only on the direction.\\
 It is easy to show by diagonalising Eq. (3) for a Hubbard square lattice that the ground state energy is the same as the bandwidth W = -8 eV = 2zt (for t = -1) where z = 4 is the coordination number. However, Eq. (1) needs to be modified for the hole motion in the $CuO_2$ plane. For as it is obvious in Fig 1b, the motion is periodic in 1D since the O sites form a ring around the Cu. An immediate question is why not use the ID hopperer?  This is not possible because the basis states are 2D. Thus we see right from the beginning of this formation why the quasi 2D nature of the $CuO_2$ plane make the bandwidth of the superconducting cuprates to be smaller than W = 2zt \cite{Belinicher1996, Hozoi2007}.\\
Eq. (1) for the ID hole motion in the $CuO_2$ plane can be expressed as
\begin{eqnarray}
&&h_{t_{pp}}\vert ii\sigma, jk\bar{\sigma}>= -t_{pp}[\vert i0\sigma, i_{(0 \pm 1)}0\bar{\sigma}> + \vert 0i\sigma, 0i_{(0 \pm 1)}\bar{\sigma}>],
\end{eqnarray}
where $L_c = 0$ and $L' = 0\pm 1$ and $t_{pp}$ is for O-O hopping.\\
Then taking Eq (4) into account in Eq. (2) will result to a 4 x 4 matrix in Eq. (3) which when diagonalisd will yield W  = -2 eV. This means the bandwidth of the non-interacting case of the $CuO_2$ plane is $75\%$ smaller than that of a normal Hubbard square. This narrow band is a key feature in the high $T_c$ superconducting cuprates because it is crucial in determining all the possible hopping and interacting parameters that will contribute to the properties of these materials and hence the model Hamiltonian \cite{Belinicher1996, Gavrichkov2005}.\\
Now if we take into account the assumption made above to retain the ZRS in the FER approach, Eq. (4) can be expressed in second quantization language as
\begin{eqnarray}
H_{t_{pd}}=-t_{pd}\biggl[\sum_{\{i\}}\sum_{<j,k>\epsilon\{i\}} d_{{i}_{r'}\sigma}^+ p_{{j}_{r'}\sigma}^+ d_{{i}_r\sigma}p_{{k}_r\sigma}+ H.C\biggr],
\end{eqnarray}
where $d^+(d)$  is the creation (annihilation) of a hole at the $Cu\ 3dx^2-y^2$ orbital and $p^+(p)$  is the creation (annihilation) of a hole at the $O\ 2p_{x}$ and $O\ 2p_{y}$ orbitals, H.C. is the Hermitian conjugate while $\{i\}$ denotes $[i,r,r',\sigma,\bar{\sigma}]$ with r =(x,y) and $r' = (y,x)$.\\
It is seen immediately that $H_{t_{pd}}$ is the kinetic operator of the t-J model \cite{Zhang1988, Hayn1993}. \\
The beauty of the HSCVA is that one can start the investigation by obtaining the bandwidth of the non-interacting case. Thereafter one can introduce the desired interactions and monitor their effects. As stated above, the on-site Coulomb interaction of of the Cu holes is considered in \cite{Zhang1988} to be very large, $U_{dd} \rightarrow \infty$. They therefore considered only the exchange interaction between the Cu and O with exchange matrix, $J_{dp}$. The physical implication of this interaction is that as the singlet state propagates, there is also an exchange of spins between the Cu and O. Thus a spin exchange distributor, $h_s$ which add and subtract one spin appropriately to the spins of the basis states will   lead to new states which are just the antisymmetric states of the basis states,
\begin{equation}
h_s\vert ii\sigma, jk\bar{\sigma}> = -J_{dp}\vert ii\bar{\sigma}, jk\sigma>,
\end{equation}
where
\begin{eqnarray}
h_s=J_{dp}\biggl[\sum_{\{i\}}\sum_{<j,k>\epsilon\{i\}} d_{{i}_{r}\sigma}^+  p_{{j}_{r}\bar\sigma}^+ d_{{i}_r\bar\sigma} p_{{k}_r\sigma}+ H.C\biggr].
\end{eqnarray}
Eq. (7) is the XY limit of the anisotropic Heisenberg exchange  interaction in second quantization language \cite{Hybersten1990}, and it  is the interacting part of the t-J model in the $CuO_2$ plane with ZRS. Taking Eq. (7) into account in Eq. (2), Eq. (3) becomes
\begin{equation}
[E_\delta{_{{L}_c L'}} + T_{{L}_c L'}+ J_{{L}_c L'}]=0.
\end{equation}
where $J = J_{pd}g_s/t_{pd}g_t$ with $g_t$ and $g_s$ being the Gutzwiller renormalization factors for the kinetic operator and superexchange interactions. The inclusion of these factors is to account for the effect of strong correlations. For in deriving Eqs. (5) and (7), the arguement in \cite{Edegger 2006} has to be invoked that since the Cu on-site Coulombic interaction is strong, $U_{dd} \longrightarrow \infty$, we will have projected fermions $d_{i\sigma} \longrightarrow X_{id\sigma} = (1 - n_{id\bar{\sigma}})$ which do not obey the usual commutation rule for free fermions. However, taking into account the Gutzwiller approximation, the renormarlized fermions becomes $X_{id\sigma} = gd_{i\sigma}$ where $g = g_t = 2x/(1 + x)$ and $g = g_s = 4/(1+x)^2$ with x denoting the level of concentration of holes. Thus the effective Hamiltonian is a combination of Eqs. (5) and (7) with the inclusion of the renormalization factors,
\begin{equation}
H_{eff} = g_tH_{tpd} + g_sh_s
\end{equation}
\section{Discussion and Conclusion}
To observe the role of the exchange interactions in comparison with Eq. (3), Eq. (8) is diagonalized at various values of $t_{pd}$ and $J_{dp}$ without the strong correlations effect for now; it is observed that the exchange interactions has a band narrowing effect \cite{Hozoi2007}. For example, the common values in the literature \cite{Thomale2008}, $t_{pd}$ = 0.4 eV and $J_{dp}$ = 0.12 eV, gives W = -1.70 eV while the experimental value of about 1 eV for the $CuO_2$ structure \cite{Ino2000} can be obtained from the unrealistic value  $J_{dp}/t_{pd}$ = 1.0. However, the inclusion of the strong correlations effect by appropriate  doping can be used to correct this defect as $W \approx$ 1 eV for $x = 0.4$ for the common values ($J_{dp}/t_{pd}$ = 0.3) and $W \approx$ 1 eV for $x = 0.3$ for $J_{dp}/t_{pd}$ = 0.2 . This is quite encouraging as it has been earlier demostrated that the Gutzwiller projected d-wave pairing state is most stable for $x \leq 0.4$ \cite{Edegger 2006} which is a fair generalization since that of the cuprates is only stable up to $x = 0.3$ generically \cite{Hague2006, Javanmard2008}. The only exception is the Y-Ba-Cu-O samples which has an additional 1D Cu-O chains that plays a role but is not taken into account in most calculations in 2D model. Thus the emphasis here is that our formulation has shown that appropraite doping of correctly chosen $CuO_2$ configuration within the t-J model may lead to results in fair agreement with experiments. It is important to point out that over the years, several sophisticated techniques have been used to investigate the inclusion of additional hopping and interaction parameters beyond the t-J model to achieve the experimental value. Though these studies have contributed to enhanced understanding, there is still no consensus on what parameters should constitute the effective Hamiltonian to study the high $T_c$ superconducting cuprates. As stated above, the reason  may be due to the starting $CuO_2$ configuration. The most probable $CuO_2$ configuration is therefore analyzed here.\\
It is assumed that the hole is created at the O site $Y_-$ as shown in Fig. (1c) where the thick spin are before exchange interactions and the dotted spin are after the exchange interactions. Therefore two hybridized dotted spins indicates a ZRS and hence is  used  here as indicative of the ZRS propagation. From  the kinetic part of Eq. (5), the hole from the $Y_-$ will hop to $X_-$ position of the O site between Plaquettes 1 and 2. It can then form a ZRS with the Cu in Plaquette  1. But due to the non-orthogonality of the plaquettes and the equidistance of the Cu-O, it can also form a triplet state with Plaquette 2 \cite{Hayn1993}. However, this is less likely because the ZRS hybridization is stronger and consequently more favoured \cite{Ogata2008}. Furthermore, the ZRS hybridization in plaquette 1 leads to spontaneous exchange of spins from the interaction part of Eq. (9). Therefore the effective hole seen by the Cu in Plaquette 2 has an opposite spin so that Eq. (9) become effective in forming new ZRS in this plaquette, resulting in a scenario in which the effective bound state forming the ZRS is propagated from one plaquette to another in the t-J model and thereby constitute a probable Cooper channel. This mode of formation and propagation of the ZRS is supported by the observation in \cite{Dagotto1994} that the critical $T_c$ of the superconducting cuprates depends on the number of $CuO_2$ planes within a short distance of each other  in the structure.\\
Now the above description looks superficially rather too simplified to explain all the intriguing properties of the high $T_c$ in the suerconducting cuprates oweing to the present sophistication of the literature which unknowingly may have become part of the complex nature of these materials. It is therefore necessary to emphasize that the mechanism proposed here do not rule out the effects of the other parameters and the influence of electron-phonon interactions from displacement of the atoms \cite{RGunnarsson2005}. For example, the spontaneous spin exchange between the O and Cu in one plaquette usually leads to virtual superexchange  between the NN Cu atoms. Interestingly, such effects have been included as Cu-Cu exchange interactions and sometimes also as constituting additional hopping term $(t')$ in attempts to account for experimental data in several studies \cite{Hozoi2007, Ogata2008}. Thus it is hoped that the formulation here gives the underlying physics of ZRS propagation in the $CuO_2$ planes which can lead to more accurate investigation  with methods that allow unbiased treatment of necessary parameters.

\acknowledgements
I acknowledge very useful discussions with M.S. Laad, M. L. Kulic, V. Yu Yushankhai and L. Hozoi. This work was partially supported by AFAHOSITECH.\\

\end{document}